# Quantitative phase and refractive index imaging of 3D objects via optical transfer function reshaping

HERVE HUGONNET,[1,2] MAHN JAE LEE,[2,3] AND YONGKEUNPARK[1,2,4,*]

[1]*Department of Physics, Korea Advanced Institute of Science and Technology (KAIST), Daejeon 34141, South Korea*
[2]*KAIST Institute for Health Science and Technology, KAIST, Daejeon 34141, South Korea*
[3]*Department of Graduate School of Medical Science and Engineering, KAIST, Daejeon 34141, South Korea*
[4]*Tomocube Inc., Daejeon, South Korea*
*\*yk.park@kaist.ac.kr*

**Abstract**

Deconvolution phase microscopy enables high-contrast visualization of transparent samples through reconstructions of their transmitted phases or refractive indexes. Herein, we propose a method to extend 2D deconvolution phase microscopy to thick 3D samples. The refractive index distribution of a sample can be obtained at a specific axial plane by measuring only four intensity images obtained under optimized illumination patterns. Also, the optical phase delay of a sample can be measured using different illumination patterns.

## 1. Introduction

Quantitative phase imaging (QPI) involves a family of microscopy techniques that enable visualization of transparent samples via images of the phase of light transmitted through the samples [1]. QPI has the advantage that quantitative measurements can be acquired without labeling. QPI is typically used to image either the phase of the light transmitted through an object in 2D or to directly reconstruct its 3D refractive index (RI) distribution. Although obtaining the 3D RI tomogram of a sample provides more information and improves the resolution compared to simple 2D phase imaging, the acquisition is often slower owing to the need to either scan the illumination angle [2-4], rotate the sample [5, 6], or translate the focal position [7].

Many QPI techniques use coherent light sources to retrieve the optical phase information based on interferometry. Although laser interferometry is a straightforward method for phase retrieval, the instruments are expensive and sensitive to mechanical vibrations as well as speckle noise [8, 9]. Recently, various QPI techniques that use incoherent light sources have been reported, including transport-of-intensity equation [10-13], quadriwave lateral-phase grating [14, 15], spatial light-interference microscopy [16, 17], gradient light-interference microscopy [18], Fourier ptychographic microscopy [19, 20], intensity diffraction tomography [21, 22], and holography via space-domain Kramers–Kronig relations [23-26].

Phase deconvolution microscopy is also based on intensity imaging and can be used to retrieve phase information from deconvolution of transmitted intensity images of a sample under different Köhler illuminations [27, 28]. The use of spatiotemporally incoherent light in phase deconvolution microscopy allows greater stability, economy and reduced speckle noise artifacts [29]. Both 2D and 3D deconvolution phase microscopy techniques have been developed. 2D phase deconvolution microscopy, also called quantitative differential phase contrast [30-35], retrieves the phase delay map by assuming a thin sample. On the other hand, 3D deconvolution microscopy, also called partially coherent optical diffraction tomography (ODT) [18, 36-39], retrieves the RI distribution using axial scanning. 2D phase deconvolution has the advantage of achieving high frame rates and being easy to implement. However, its reconstruction is limited to thin samples and does not provide direct RI measurements.

Herein, we develop a method to obtain either the RI distribution or optical phase delay images of thick samples at specific axial planes without requiring axial scanning. By measuring only four intensity images of a sample with optimized illumination patterns, the RI distribution of the sample at a specific axial plane can be quantitatively and precisely retrieved (Fig. 1). Moreover, the 2D optical phase delay image can be calculated in the same way with differently optimized illumination patterns. We present the principle, optical instrumentation, and validation with the microspheres and unlabeled cancer cells.

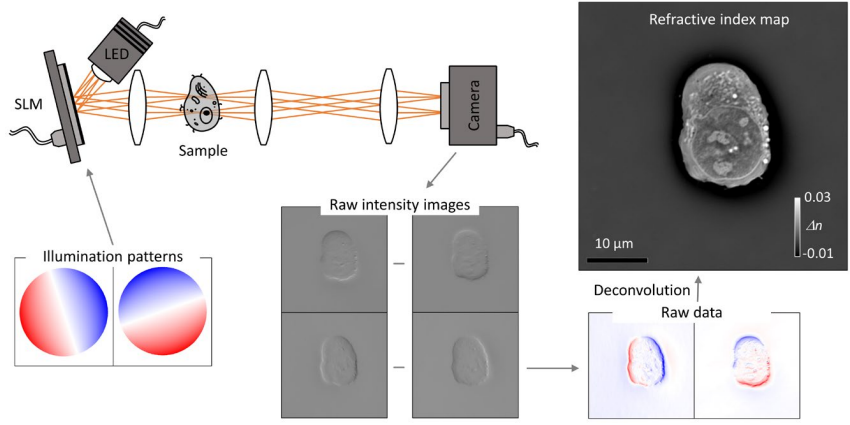

Fig. 1. Principle of the proposed method, where the illumination is controlled in the Fourier plane and then the transmitted intensity is deconvolved to obtain the refractive index (RI). LED: light-emitting diode, SLM: spatial light modulator.

## 2. Methods

### 2.1 Deconvolution phase microscopy

In deconvolution phase microscopy [36], a sample is illuminated by an incoherent beam, whose intensity pattern is controlled in the Fourier plane, to retrieve the RI distribution. The light intensity transmitted through the sample is then recorded at various focal planes and deconvolved to obtain the RI. The relative variation between the intensity transmitted through the object and illumination $S = (I_{out} - I_{in})/I_{in}$ is expressed as the convolution of the sample's scattering potential $F(x,y,z) = \frac{1}{4\pi} k^2 \left[ n^2(x,y,z)/n_m^2 - 1 \right]$ with a point-spread function (PSF), where $n$ and $n_m$ are the RIs of the sample and surrounding medium, respectively. The PSFs are different for the real and imaginary parts of the scattering potential and are respectively called the phase optical transfer function (OTF) $H_P$ and amplitude OTF $H_A$:

$$\widetilde{S} = \widetilde{H_A} \widetilde{F_{img}} + \widetilde{H_P} \widetilde{F_{real}}, \tag{1}$$

where ~ indicates the Fourier transform and $F = F_{real} + i \cdot F_{img}$.

The OTFs are computed by convolution between the green function and illumination intensity distribution in the Fourier space [28, 37, 40] as follows:

$$\begin{cases} H_P(\mathbf{k}) = -128\pi^4 \cdot \mathrm{Im}\left( \widetilde{G^* \cdot \widetilde{G k_z \rho}} \right) \\ H_A(\mathbf{k}) = -128\pi^4 \cdot \mathrm{Re}\left( \widetilde{G^* \cdot \widetilde{G k_z \rho}} \right) \end{cases}, \tag{2}$$

where Im() describes the imaginary part, Re() is the real part, $G(x,y,z) = \frac{i}{8\pi^2} \iint_{\sqrt{k_x^2+k_y^2}<NA} \frac{e^{ik_x x + ik_y y + ik_z z}}{k_z} dk_x dk_y$ is the green function truncated to the optical system numerical aperture, $k_z = \sqrt{k^2 - k_x^2 - k_y^2}$ is the axial wavenumber, $k$ is the wave number, and $\rho(k_x, k_y)$ is the illumination intensity distribution in the Fourier plane. When several illumination patterns $\rho_1, \rho_2, ...$ are used, the relation between the raw data and scattering potential can be expressed [28] as

$$\widetilde{\overline{\overline{S}}} = \begin{pmatrix} \widetilde{S_1} \\ \widetilde{S_2} \\ ... \end{pmatrix} = \begin{bmatrix} \widetilde{H_{A1}} & \widetilde{H_{P1}} \\ \widetilde{H_{A2}} & \widetilde{H_{P2}} \\ ... & ... \end{bmatrix} \begin{pmatrix} \widetilde{F_{img}} \\ \widetilde{F_{real}} \end{pmatrix} = \widetilde{\mathbf{H}} \widetilde{\overline{\overline{F}}}, \tag{3}$$

with the numerical index referring to the illumination pattern number. The deconvolution is then achieved by computing the pseudoinverse of the OTF matrix [37] as

$$\widetilde{\overline{\overline{F}}} = \left( \widetilde{\mathbf{H}} \widetilde{\mathbf{H}}^T + \alpha \mathbf{I} \right)^{-1} \widetilde{\mathbf{H}}^T \widetilde{\overline{S}}, \tag{4}$$

where α is a regularization term to avoid noise amplification. The imaging quality depends directly on the OTF. Several works on both 2D and 3D imaging highlight the importance of a homogeneous OTF to maximize imaging quality [31, 37, 41]. The illumination intensity distribution in the Fourier plane is then optimized to maximize the OTF homogeneity.

The homogeneity of the transfer function can be evaluated using the OTF density (OTFD) [42] given by $\text{OTFD}(\mathbf{k}) = \sqrt{\sum_{j=1}^{N}\left[\widetilde{H}_{jP}^{-1}(\mathbf{k})\right]^2}$. Plural efficient patterns for self-interfering ODT (PEPSI ODT) [37] uses this metric to obtain a homogeneous OTF that enables artifact-free 3D reconstruction of the RI; therefore, PEPSI ODT is used as the standard for comparison with our method in this work.

*2.2 2D imaging of 3D objects*

Deconvolution phase microscopy has also been widely used for 2D imaging [31, 32, 37]. By assuming that a sample is thin and in focus $F = F \cdot \delta(z)$, the spectrum of the sample scattering potential is constant in the vertical direction $\widetilde{F}(k_x, k_y, k_z) = \widetilde{F}(k_x, k_y, 0)$. Because the spectrum of the sample is constant in the vertical direction, the problem can be simplified to two dimensions by defining the OTF as the incremental sum of the 3D OTF along the vertical direction $\widetilde{H_{P2D}}(k_x, k_y) = \int_{-\infty}^{\infty} \widetilde{H_P}(k_x, k_y, k_z) dk_z$ and imaging the intensity transmitted through the sample only at the plane of best focus [33]. However, if the sample is partially out of focus or thicker than the optical sectioning of the setup, imaging artifacts may occur.

Here, we propose another approach to 2D phase deconvolution microscopy, instead of supposing that the sample spectrum is constant in the vertical direction. We design the illumination distribution in the Fourier plane so that the OTF is constant in the vertical direction:

$$\begin{cases} \widetilde{H_P}(k_x, k_y, k_z) = \widetilde{H_P}(k_x, k_y, 0) & \text{if } |k_z| < k_{z\max} \\ \widetilde{H_P}(k_x, k_y, k_z) = 0 & \text{if } |k_z| > k_{z\max} \end{cases}. \quad (5)$$

More precisely, the OTF must be constant in the vertical direction within the deconvolution bandwidth $k_{zmax}$; then, the 2D deconvolution process is similar, but the method can be used to image thicker or out-of-focus samples. While the OTF must be constant in the vertical direction within the deconvolution bandwidth, the deconvolution bandwidth itself can be chosen to increase or decrease the depth of field. As proof of concept, we optimized two illumination schemes, with the first one having a shallow depth of field by maximizing the deconvolution bandwidth to compute the sample RI at a specific axial plane, which we named single section RI deconvolution (SISRID). The second illumination scheme was optimized for an extended depth of field by minimizing the deconvolution bandwidth to compute the optical path length or phase transmitted through an object, which we named the single section optical path deconvolution (SISOPD).

To obtain an OTF meeting these criteria, we used the optimization pipeline described in Ref. [31] based on simulated annealing [37] to design the illumination intensity distribution in the Fourier plane. The minimized cost functions were chosen as follows:

$$\begin{cases} \text{cost}_{SISRID} = \alpha_1 \frac{\sum_{\mathbf{k}} |\nabla \text{OTFD}(\mathbf{k})|}{\left(\sum_{\mathbf{k}} |\text{OTFD}(\mathbf{k})|\right)^{1.01}} + \beta_1 \frac{\sum_{\mathbf{k}} |\nabla_{k_z} H_P(\mathbf{k})|}{\left(\sum_{\mathbf{k}} |H_P(\mathbf{k})|\right)^{1.01}} \\ \text{cost}_{SISOPD} = \alpha_2 \frac{\sum_{\mathbf{k}} |\nabla \text{OTFD}(\mathbf{k})|}{\left(\sum_{\mathbf{k}} |\text{OTFD}(\mathbf{k})|\right)} + \beta_2 \frac{\sum_{\mathbf{k}} |\nabla_{k_z} H_P(\mathbf{k})|}{\left(\sum_{\mathbf{k}} |H_P(\mathbf{k})|\right)} + \gamma_2 \frac{\sum_{k_x k_y} |\text{OTFD}(k_x, k_y, 0)|}{\left(\sum_{\mathbf{k}} |\text{OTFD}(\mathbf{k})|\right)} \end{cases}. \quad (6)$$

The constant weights $\alpha_1 = 10, \beta_1 = 10, \alpha_2 = 5, \beta_2 = 2.5, \gamma_2 = 50$ were chosen such that each term of the cost function had a similar value when applied to a random pattern. Each term in the cost function has a specific role: the first term ensures a homogenous OTF in all directions and is the same term as that used in the PEPSI ODT [43] to avoid directional artifacts. The second term ensures that the phase transfer function is constant in the vertical direction, which is the necessary condition for 2D deconvolution. Finally, the third term used only for SISOPD minimizes the deconvolution bandwidth.

Notably, both the phase and amplitude transfer functions could not be optimized as constants in the vertical direction simultaneously, so we assumed that the sample showed no absorption and set $H_A = 0$ during reconstruction. To minimize the effects of light absorption by the sample, we added a central symmetry constraint to the Fourier-plane illumination intensity distribution to minimize the magnitude of the amplitude transfer function. The positive and negative parts of the illumination pattern were imaged sequentially, and the transmitted intensity images through the sample were subtracted to obtain the raw data.

As such a total of four intensity measurements were used however the number of measurements could easily be reduced to three by using the same illumination to obtain the negative part of both pattern and to two measurement if not imposing central symmetry on the patterns. However, note that such strategies might increase measurement noise.

The obtained optimized illuminations and corresponding OTFs are plotted in Fig. 2. As required, they show radially uniform OTF densities and constant OTF values in the vertical direction.

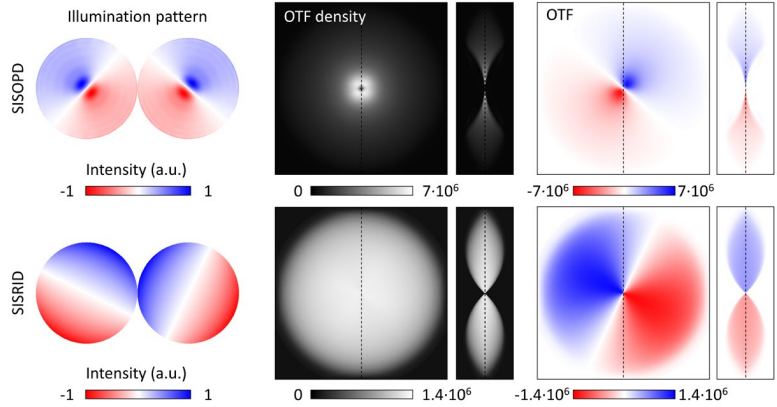

Fig. 2. Illumination patterns, OTF densities, and OTFs obtained after optimization.

### 2.2 2D optical setup

For demonstration purposes, we built a custom optical setup with a light-emitting diode (LED) illumination source. The illumination intensity distribution was controlled in the Fourier plane using a spatial light modulator (SLM; Forth dimension display M150) and was then projected onto a sample through a water-immersion condenser lens (Olympus UPLSAPO60XW, NA 1.2). The diffracted light from the sample was collected using a water-immersion objective lens (Olympus UPLSAPO60XW, NA 1.2), projected onto the image plane, and then measured using a CMOS camera (Flir ORX-10G-71S7M-C, 250 fps at 2×2 binning).

### 2.4 Sample preparation

To validate the proposed method, we applied our imaging technique to biological samples. Live cells from the acute myeloid leukemia (AML) cell line MV4-11 were cultured in an RPMI-1640 medium (Gibco®, CA, USA) supplemented with 10% fetal bovine serum (Gibco®, CA, USA) and 1% penicillin/streptomycin (Sigma Aldrich, MO, USA) in humidified temperature containing 5% carbon dioxide ($CO_2$) at 37 °C. The cells were reloaded on imaging culture dishes (Tomodish, Tomocube Inc., Daejeon, Korea) and covered with a coverslip before imaging.

## 3. Results

### 3.1 Benchmarking with synthetic samples

We first imaged polymethylmethacrylate ($n$ = 1.491) bead clusters in an ultraviolet curing resin ($n_m$ = 1.48), and the results are shown in Fig. 3. The SISRID 2D section RI images precisely matched the 3D tomogram obtained via PEPSI ODT without requiring axial scanning of the sample. Furthermore, the SISOPD sections could visualize all four beads in a cluster even when varying focus positions, thus demonstrating the increase in depth of field (Fig. 3(a)). The measured RI values and optical path lengths were consistent with manufacturer specifications.

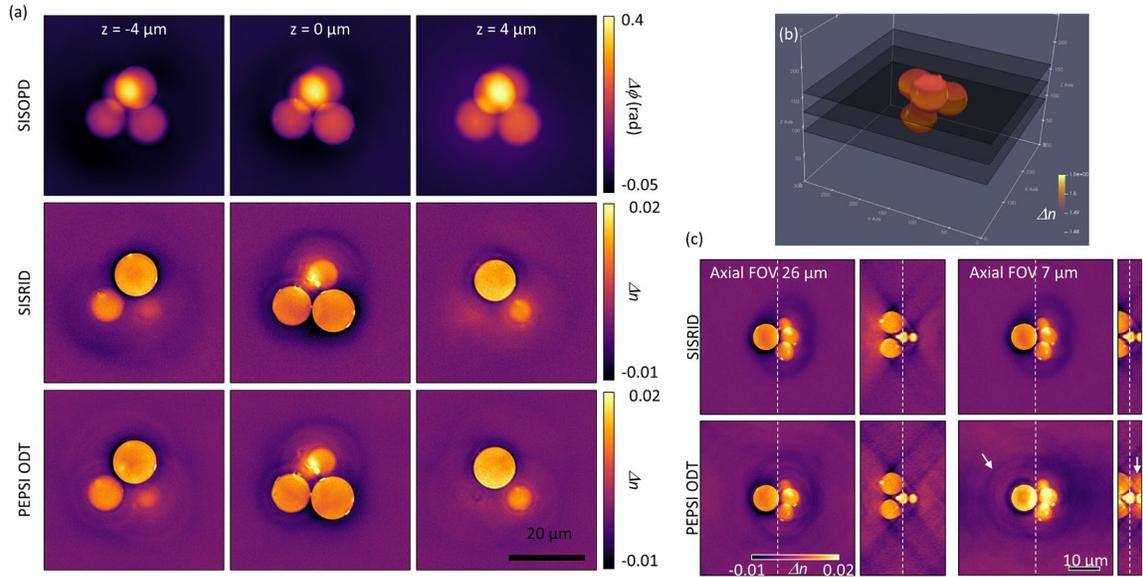

Fig. 3. (a) Imaging results of beads using SISOPD, SISRID, and PEPSI ODT; (b) 3D rendering of RI tomogram obtained via SISRID; (c) 3D imaging of beads by SISRID and PEPSI ODT with reduced fields of view.

By axially scanning the sample while acquiring multiple 2D images, our proposed method can also be used to perform 3D imaging (Fig. 3(c)). Since every axial position in the sample is reconstructed independently, our method alleviates the boundary artifacts in 3D PEPSI ODT, thereby enabling imaging of only the volume of interest. For example, artifacts are observed when reducing the axial field of view in PEPSI ODT but not in the SISRID (white arrows in Fig. 3(c)).

*3.2 Results with biological samples*

To verify the applicability of the proposed method, we repeated the experiments on biological samples (Fig. 4). Both the RI distributions and OPD images of unlabeled live cancer cells from the MV4-11 cell line were measured. The SISRID visualizes the sample with good optical sectioning displaying different features at each focus position, whereas the SISOPD displays similar cell features independent of the focus position to demonstrate the large depth of focus.

The high acquisition speed (~50 Hz) of our method enables live views of the biological samples and imaging of rapid dynamics, which could be further increased using a color camera [37]. The high acquisition speed can be further exploited to obtain higher sensitivity through averaging. Fig. 4(d) shows live cells imaged with high sensitivity by averaging 50 frames. Compared with the images before averaging, as in Fig. 4(c), the noise reduction enables better visualization of the cell organelles.

The sensitivity of the setup was computed by evaluating the standard deviation of the RI in a region of 100×100 pixels in the absence of a cell. The standard deviation was reduced from $6.60\times10^{-4}$ to $1.28\times10^{-4}$ by averaging. This 5-fold noise reduction is consistent with the $\sqrt{50}\simeq 7$ fold reduction expected from averaging 50 images with shoot noise.

The reconstructed RI distribution of the cells clearly shows various subcellular structures (Fig. 4(d)). For example, the top image shows the extrusion of the nuclear membrane (yellow arrow) and nucleoli (red arrow). The tubular-shaped subcellular organelles appear to be mitochondria (white arrow in the middle image). The round-shaped organelles with high RI values are lipid droplets [44]. These results suggest that structural information on individual cells and subcellular organelles can be investigated in an unlabeled live state using the proposed approach, whereas conventional techniques such as scanning electron microscopy only show static images of dead cells [45, 46].

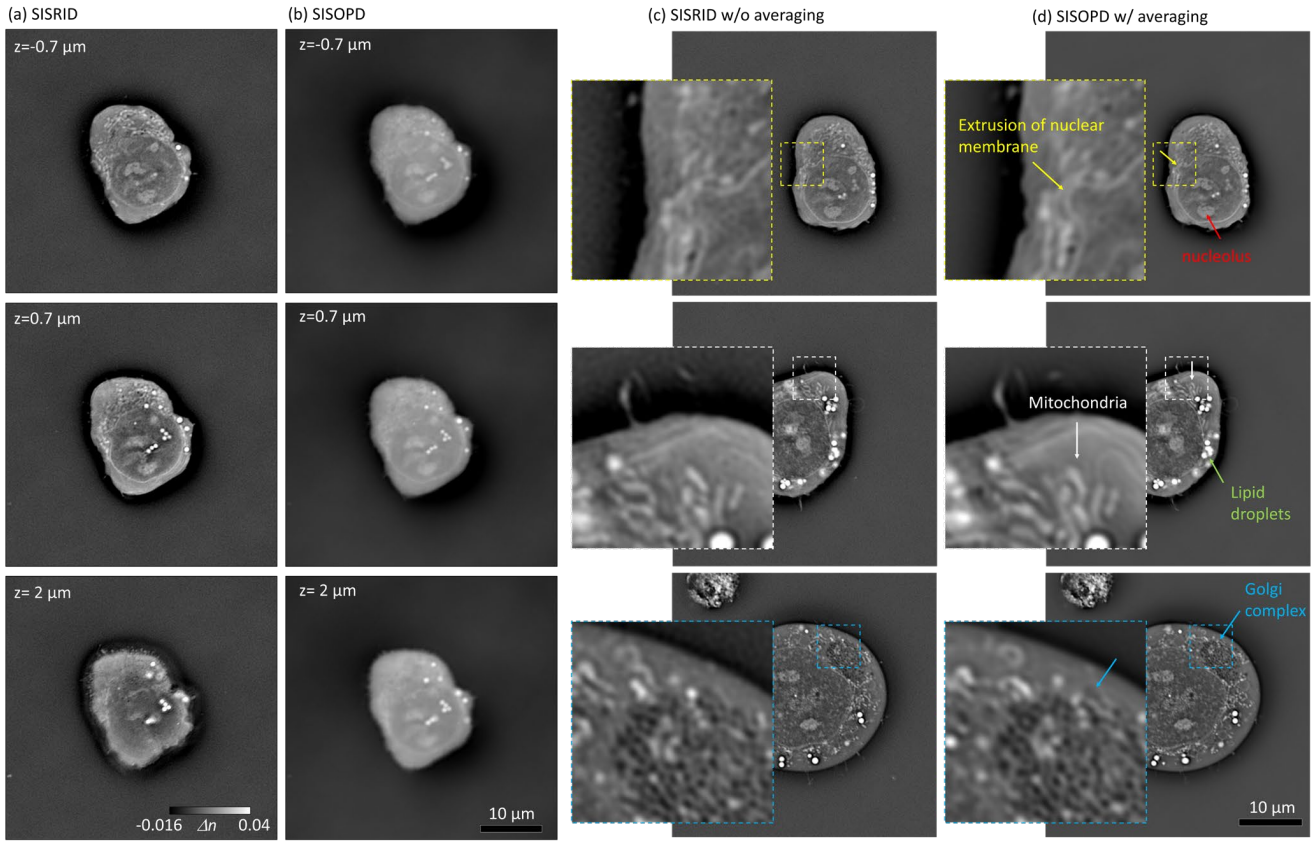

Fig. 4. Imaging of biological cells by (a) SISRID at different focal planes, (b) SISOPD at different focal planes, and SISRID (c) without and (d) with averaging.

## 4. Discussion

We proposed a methodology to remove the depth limitation and enable imaging of thick or partially out-of-focus samples. Furthermore, the depth of focus could be adjusted using different illumination patterns. The imaging results were verified against those of 3D deconvolution phase microscopy (PEPSI ODT) and observed to display similar fidelities for both synthetic and biological samples. Because only four intensity measurements are required without sample scanning to obtain the RI image, the proposed method can be used to investigate fast phenomena and provide live video-rate previews of the RIs. This method can also be used to investigate slight variations in the RI by increasing the signal-to-noise ratio by combining multiple RI images. When our method is combined with axial scanning 3D imaging is achieved, the proposed method removes wrap-around artifacts present in the 3D deconvolution phase microscopy, thereby enabling a more flexible field of view.

## 5. Acknowledgements

This work was supported by KAIST UP program, BK21+ program, Tomocube, National Research Foundation of Korea (2015R1A3A2066550), and Institute of Information & communications Technology Planning & Evaluation (IITP; 2021-0-00745) grant funded by the Korea government (MSIT)

## 6. Disclosures.

H.H, M.L and Y.P. have financial interests in Tomocube Inc., a company that commercializes optical diffraction tomography and quantitative phase-imaging instruments, and is one of the sponsors of the work.